\documentclass{article}

\usepackage{PRIMEarxiv}

\usepackage[utf8]{inputenc} 
\usepackage[T1]{fontenc}    
\usepackage{hyperref}       
\usepackage{url}            
\usepackage{booktabs}       
\usepackage{amsfonts}       
\usepackage{nicefrac}       
\usepackage{microtype}      
\usepackage{lipsum}
\usepackage{fancyhdr}       
\usepackage{graphicx}       
\graphicspath{{media/}}     
\usepackage{subfig}
\usepackage{algorithm}
\usepackage{algpseudocode}
\usepackage{physics}
\usepackage{tikz}
\usetikzlibrary{quantikz}

\title{Implementing An Artificial Quantum Perceptron}

\author{
  Ashutosh ~Hathidara\thanks{The work was done as part of Masters degree at Indiana University.}\\
  SAP AI \\
  \texttt{ashutosh.hathidara@sap.com} \\
   \And
  Lalit Pandey \\
  Indiana University \\
  \texttt{lpandey@iu.edu} \\
}

\begin{document}
\maketitle

\begin{abstract}
A Perceptron is a fundamental building block of a neural network. The flexibility and scalability of perceptron make it ubiquitous in building intelligent systems. Studies have shown the efficacy of a single neuron in making intelligent decisions. Here, we examined and compared two perceptrons with distinct mechanisms, and developed a quantum version of one of those perceptrons. As a part of this modeling, we implemented the quantum circuit for an artificial perception, generated a dataset, and simulated the training. Through these experiments, we show that there is an exponential growth advantage and test different qubit versions. Our findings show that this quantum model of an individual perceptron can be used as a pattern classifier. For the second type of model, we provide an understanding to design and simulate a spike-dependent quantum perceptron. Our code is available at \url{https://github.com/ashutosh1919/quantum-perceptron}
\end{abstract}


\section{Introduction}
A perceptron is an artificial unit of an intelligent system capable of making decisions. This artificial unit is inspired by the biological neurons found in the human brain. The human brain has a network of billions of neurons connected to each other. This connectivity leads to the formation of a deep network. Thus, a perceptron is used as a fundamental building block in deep learning systems. In classical computing, these perceptrons have two states, 0 and 1. When the input of the perceptron is sufficient enough to generate an output over the threshold limit, the perceptron is said to be in `ON' or 1 state. On the other hand, if the output of the perceptron is less than its threshold value, then it is in `OFF' or 0 state \cite{paper1}. 

Decades of research in the field of classical deep learning have given rise to state-of-the-art systems \cite{paper21} \cite{paper22} that mimic human-level intelligence. Drawing from recent research that suggests the role of quantum entanglement in consciousness, there has been growing interest in exploring the potential of quantum computing to advance artificial intelligence. However, despite this progress, there remains a gap when implementing quantum algorithms in AI. In this study, we aim to bridge this gap by implementing a quantum model of a perceptron. Here, we review the available literature \cite{quantum_perceptron} and implement the quantum circuit using Qiskit quantum simulator \cite{qiskit} to simulate the training of a single perception.

Almost every advanced deep learning system has artificial neurons as the fundamental building block. Inspired by the success in the classical machine learning field, we attempt to implement a quantum version of a perceptron that mimics the properties of a classical perceptron but has the benefits of a quantum system and obeys the rules of quantum mechanics.

Previous works like \cite{quantum_perceptron} introduce a novel architecture and quantum algorithm to design a quantum version of a perceptron. We examine the algorithm and simulate it to test the efficacy of the quantum algorithm. For the implementation, we use QisKit quantum simulation tool and construct quantum gates as specified in the algorithm. We then develop an end-to-end pipeline to generate datasets, initialize weights,  train the perceptron, and simulate the probability behavior as discussed in \cite{quantum_perceptron}. Following the training process, we conduct a comprehensive analysis to confirm the trained perceptron's ability to accurately classify patterns.

\begin{figure}[htp]
    \centering
    \includegraphics[width=6cm]{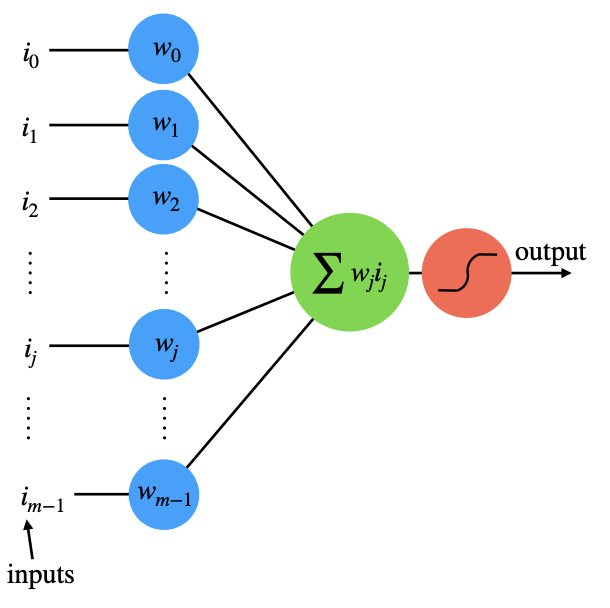}
    \caption{A classical perceptron used in deep learning systems. The perceptron takes multiple input values $\{i_0, i_1, \hdots, i_{n-1}\}$. Internally, it initializes random weight values $\{w_0, w_1, \hdots, w_{n-1}\}$ corresponding to each of the input values. The perceptron computes the dot product of the input and weight vector i.e. $\vec{i} \cdot \vec{w} = \sum_{j=0}^{n-1} i_jw_j$. This dot product result is passed through a non-linear sigmoid \cite{sigmoid_fn} function which computes the probability. This probability can be used to compute the loss using the supervised label. The computed loss can then be used to train the perceptron by backpropagating gradients \cite{backpropagation} and updating the weights.}
    \label{fig:base_perceptron}
\end{figure}

\section{Related Work}
The concept of a perceptron was first introduced in \cite{mcculloch_pitts_perceptron}, which presented the classical mathematical framework for utilizing a perceptron as a supervised data classifier. Numerous successful examples have demonstrated the effective application of this mathematical principle in real-world scenarios.

In 2013, Lloyed et. al. \cite{lloyd2013quantum} introduced a theoretical notion of quantum perceptron for supervised and unsupervised learning. Such perceptrons require generalized values and use qRAM \cite{qram} to store values. This study contributes to the theoretical literature of quantum computing. In 2014, Schuld et. al. \cite{Schuld_2015} introduced the concept of simulating perceptrons using tools. They used the same simulation tools used in \cite{quantum_perceptron} to implement the quantum circuit of a perceptron. The terminology and the approach are similar too. However, \cite{Schuld_2015} utilizes QFT to create intermediate oracle circuits to prepare the input and weight states which operates on an exponential number of gates. On the other hand, \cite{quantum_perceptron} make use of hypergraph states to construct these oracles. This allows them to operate with a polynomial number of quantum gates. The most recent classical deep learning models, as described in \cite{dl_overview}, utilize bias vectors in addition to weight vectors for their perceptrons. As implementing perceptron algorithms in the quantum field is a relatively new concept, we omit the bias vector and exclusively focus on training the weight vector.

\section{Methods}

\subsection{Architecture}
Unlike a classical perceptron, a quantum perceptron has quantum gates that prepare the inputs and weights for the system to process. Unitary transformation functions are used to pre-process the input and weight vectors. A Unitary transformation function, also known as an Oracle, houses quantum gates which act upon the input vectors to perform operations such as phase shift, imposing superposition, entanglement, etc. Akin to classical neurons, a quantum perceptron takes an input vector and a weight vector and outputs a probability of the outcome.

\begin{figure}[htp]
    \centering
    \includegraphics[width=7cm]{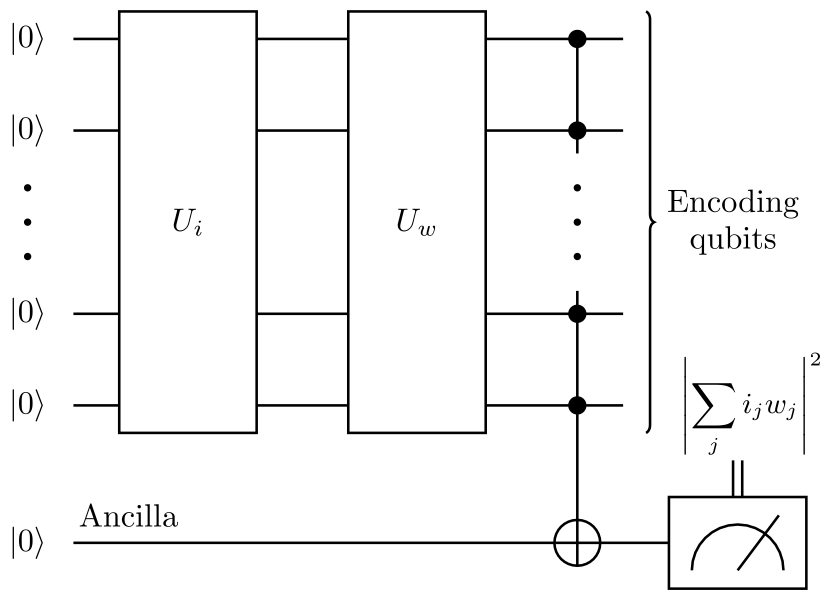}
    \caption{A Quantum Version of Perceptron.}
    \label{fig:perceptron_arch}
\end{figure}

\begin{equation}\label{eqn:input_transform}
    \ket{\psi_i} = \frac{1}{\sqrt{m}}\sum_{j=0}^{m-1} i_j \ket{j}
\end{equation}

\begin{equation}\label{eqn:weight_transform}
    \ket{\psi_w} = \frac{1}{\sqrt{m}}\sum_{j=0}^{m-1} w_j \ket{j}
\end{equation}

Figure \ref{fig:perceptron_arch} illustrates the internal structure of a perceptron architecture. Two Unitary transformation functions namely, $U_i$ and $U_f$, are used to perform quantum operations. The input vector is transformed into an input state by applying the $U_i$ function as shown in equation \ref{eqn:input_transform}, while the $U_f$ function transforms the weight vector into a weighted state as shown in equation \ref{eqn:weight_transform}. After applying the transformation functions, the dot product is calculated between the input and the weight state ($\braket{\psi_w}{\psi_i}$). This entire series of operations are carried out until the model converges and we obtain the optimal weight.

\subsection{Dataset Generation}

We used the same quantum perceptron to generate the dataset consisting of value-label pairs. Following Mcculloch et. al. \cite{paper1}, we replaced all the classical bits containing 1 with -1 and 0 bits with 1. For instance, if the input value is 12, then the transition from classical to quantum vector will look as $12 \rightarrow [1, 1, 0, 0] \rightarrow [-1, -1, 1, 1] $. 

\begin{figure}[htp]
    \centering
    \includegraphics[width=6cm]{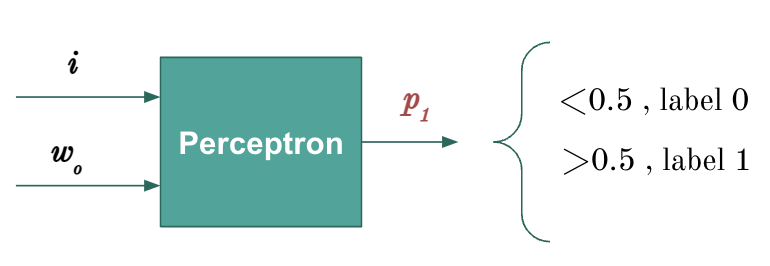}
    \caption{Generating dataset using single perceptron}
    \label{fig:perceptron_data_gen}
\end{figure}

The overall implementation of dataset generation is described in algorithm \ref{alg:data_gen}. The algorithm was tested using varying numbers of qubits, resulting in 16 possible input values when using 2 qubits and $2^{16}$ possible input values when using 4 qubits. 

\begin{figure}[htb]
  \centering
  \begin{minipage}{.9\linewidth}
\begin{algorithm}[H]

\caption{Data Generation}\label{alg:data_gen}
\begin{algorithmic}[1]
\Require Optimal weight $w_o$, Number of qubits $n$, Number of iterations $N$
\State $data \gets \{\}$ \Comment{Initializing empty list}
\State $p \gets Perceptron(n)$ \Comment{Initializing perceptron}
\For{$i \in [0, 2^{2^n} - 1]$}
\State $p.input \gets i$
\State $p.weight \gets w_0$
\State $p_1 \gets p.measure(N)$ \Comment{Probability of measuring 1}
\If{$p_1 < 0.5$}
    \State $data.add((i, 0))$ \Comment{Assigning label 0}
\ElsIf{$p_1 \geq 0.5$}
    \State $data.add((i, 1))$ \Comment{Assigning label 1}
\EndIf
\EndFor
\end{algorithmic}
\end{algorithm}
\end{minipage}
\end{figure}

A neural network requires a dataset to operate upon and to update the network's parameters. To generate the dataset, first, we take a fixed optimal weight $w_0 = 626$ as shown in figure \ref{fig:perceptron_data_gen}. Second, we passed sequential input values and weight $w_o$ to the perceptron. Finally, we compute the output  probability and based on that label the data. If the probability was less than 0.5, the input value was classified as 0, and if it was 0.5 or greater, the input value was classified as 1. The weight was constant and did not update throughout the data collection process. This approach is similar to supervised learning in the case of classical deep learning systems.

\subsection{Training}

Classical deep learning systems need an enormous amount of training to achieve convergence. In contrast, quantum computing offers the advantage of rapidly converging models. The quantum perceptron training is described in the algorithm \ref{alg:training}.

\begin{figure}[htb]
  \centering
  \begin{minipage}{.9\linewidth}
\begin{algorithm}[H]
\caption{Training Perceptron}\label{alg:training}
\begin{algorithmic}[1]
\Require Optimal weight $w_o$, Number of qubits $n$, Number of iterations $N$, $data$
\State $w_t \gets U(0, 2^{2^n} - 1)$ \Comment{Randomly initialize weight for training}
\State $p \gets Perceptron(n)$ \Comment{Initializing perceptron}
\For{$i,l \gets data$}
\State $p.input \gets i$
\State $p.weight \gets w_t$
\State $p_1 \gets p.measure(N)$ \Comment{Probability of measuring 1}
\If{$p_1 < 0.5$ and $l = 1$}
    \State FLIP-NON-MATCHING-BITS($w_t, i$) \Comment{Flip non-matching bits of $w_t$ w.r.t $i$}
\ElsIf{$p_1 \geq 0.5$ and $l = 0$}
    \State FLIP-MATCHING-BITS($w_t, i$) \Comment{Flip matching bits of $w_t$ w.r.t $i$}
\EndIf
\State converged if $w_t = w_o$
\EndFor
\end{algorithmic}
\end{algorithm}
\end{minipage}
\end{figure}

During the training phase, each perceptron is initialized with a random weight which is updated after each training iteration. Here, for a system with 4 qubits, we initialize a random weight $w_t$. The goal of training the perceptron is to update its weights, such that it can correctly classify the input values as per their labels. In case when the misprediction happens, we need to penalize the loss such that the weights are updated. Here, we have two cases of misprediction. Below, we describe the details to handle the misprediction to update weights.

\paragraph{Case 1:} Predicted label = 0, Actual label = 1.
In this case, we first find the number of non-matching bits between the input and weight sequence. Next, we multiply the learning rate by the number of non-matching bits and round down to obtain a product. Finally, we randomly flip the resulting number (product obtained in the above step) of bits in the weight, bringing it closer to the input sequence and facilitating faster convergence of the model.
\paragraph{Case 2:} Predicted label = 1, Actual label = 0.
In this case, instead of finding non-matching bits, we look for the matching bits between the input and weight sequence. The rest of the steps remain the same as in case 1. \\

The weight of the perceptron is updated after each training iteration (epoch) based on the above two cases. Note that we do not need to update the weights when the prediction is correct since the loss in such cases would be zero. Finally, we check if $w_t = w_o$ and stop the training if satisfied.

\section{Results}
\begin{figure}[htp]
    \centering
    \includegraphics[width=7cm]{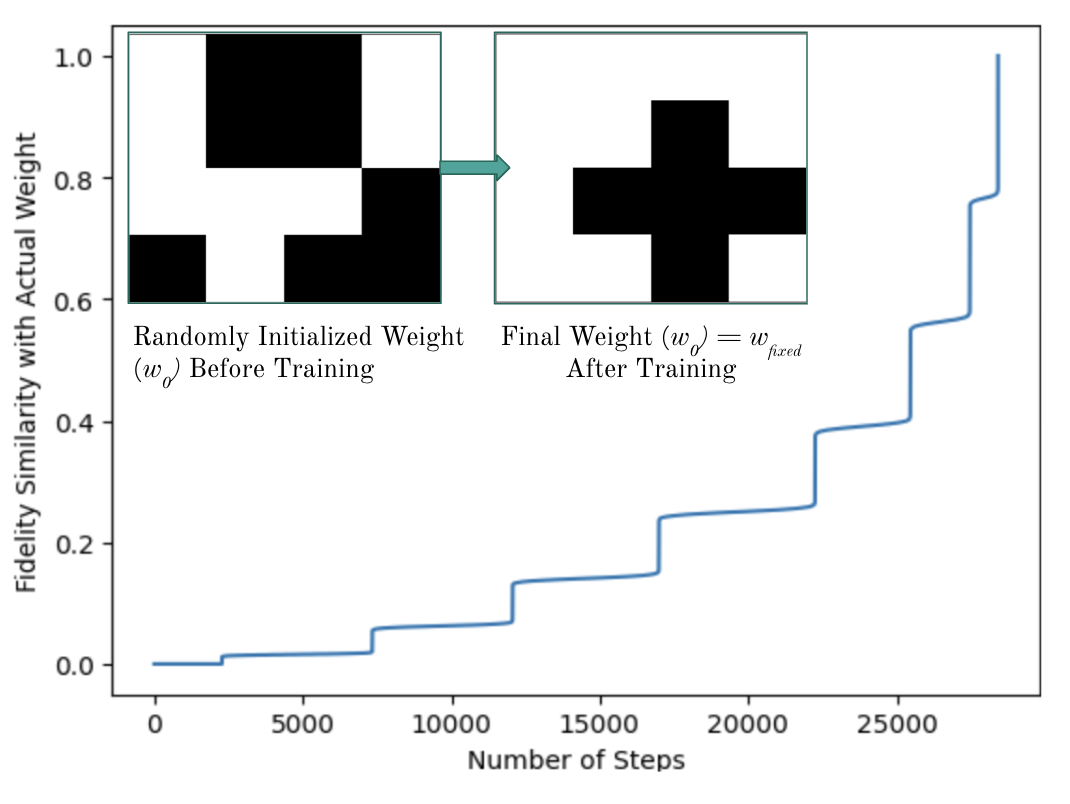}
    \caption{Training procedure for the generated data}
    \label{fig:training_img}
\end{figure}

\paragraph{Pattern Classification:} We trained a quantum perceptron and visualized its optimal weights after training. Figure \ref{fig:training_img} shows the training steps and the transformation of randomly initialized weight into a complete pattern. Through our experiments, we found that a single quantum perceptron can successfully classify simple patterns of horizontal and vertical lines. Here, we report one such pattern after training the perceptron.

\paragraph{Faster Convergence:} Compared to classical deep learning systems, a quantum perceptron can achieve optimal performance faster and has the ability to terminate training once the optimal weight has been reached. We found that a four-qubit system converged and reached the optimal weight before the training was completed.

\begin{figure}[htp]
    \centering
    \subfloat[\centering Simulation for 2 qubit system]{{\includegraphics[width=5cm]{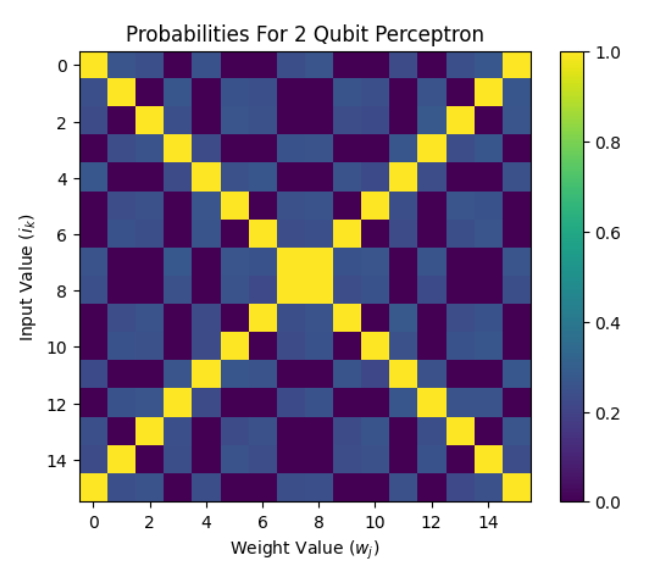}}}%
    \qquad
    \subfloat[\centering Simulation for 3 qubit system]{{\includegraphics[width=5cm]{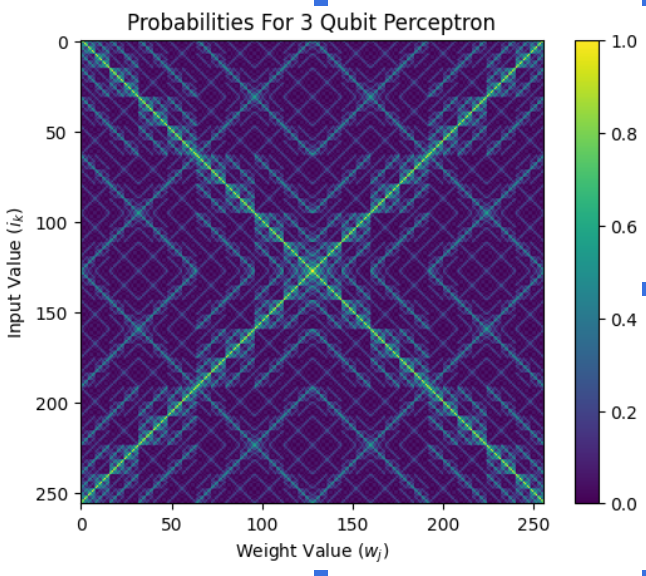} }}%
    \caption{Simulation of perceptron on all combinations of input and weight values}%
    \label{fig:simulations}%
\end{figure}

\paragraph{Identical Input and Weight:} Finally, we only get a probability of 1 when the input and the weight have the same value. The geometrical patterns in figure \ref{fig:simulations} denote the perceptron probability for all combinations of input and weight values.

\section{Conclusion and Future Work}
We implemented a quantum version of a perceptron and tested the algorithm's efficacy. Upon analysis, a single perceptron was able to classify patterns after training. The results suggest that a quantum perceptron converges faster than a classical perceptron. This faster convergence highlights the parallel processing of the inputs present in the superposition states. One of the limitations of this work is the use of a single perceptron to design a classifier. Another limitation is the absence of bias vectors in addition to the weight vectors in the training process. We also confine the input values (only -1 and 1) when training the perceptron. Future work will focus on designing and implementing an advanced network with more interconnected perceptrons. This will lead to the development of an advanced quantum network for classification purposes.

\section{Acknowledgment}
We thank Dr. Mohsen Heidari, professor at Indiana University Bloomington, for being our instructor, guiding us throughout the project, and thereby supporting our work.

\subsection{Conflict of Interest}
The authors declare that they have no conflict of interest.

\subsection{Funding}
The authors received no financial support for the research, authorship, and/or publication of this article.

\bibliographystyle{unsrt}  
\bibliography{references}

\end{document}